\documentclass[twocolumn]{IEEEtran}

\usepackage[colorlinks,urlcolor=blue,linkcolor=blue,citecolor=blue]{hyperref}

\usepackage{color,array,amsthm}

\usepackage{graphicx}
\usepackage{setspace}

\usepackage{graphicx}
\usepackage{amsmath}
\usepackage{mathrsfs}  % $\mathscr{text}$
\newtheorem{remark}{Remark}  % for remark
\newtheorem{notation}{Notation}  % for remark
\usepackage{subfigure}
\usepackage{multirow}  % to combine multiple row
\usepackage{diagbox}
\usepackage{multicol}  % to dan lan

\usepackage{stfloats}

\usepackage{hyperref}

\begin{document}

\title{Parameter Identification of a PN-Guided Incoming Missile Using an Improved Multiple-Model Mechanism}

\author{Yinhan Wang\thanks{Yinhan Wang is with Beijing Institute of Technology, Beijing, CO 100081, China},
Jiang Wang\thanks{Jiang Wang is with Beijing Institute of Technology, Beijing, CO 100081, China},
Shipeng Fan\textsuperscript{*} 
\thanks{Shipeng Fan is with Beijing Institute of Technology, Beijing, CO 100081, China}
\thanks{\textsuperscript{*}Corresponding Author: Shipeng Fan, Email: fspzxm@sina.com}}

%\thanks{corresponding author: Shipeng Fan}/

%\thanks{Beijing Institute of Technology, Beijing 100081, People's Republic of China}

% make the title area
%\begin{spacing}{1.0}
\maketitle
%\end{spacing}

%\begin{spacing}{1.0}
\begin{abstract}
An active defense against an incoming missile requires information of it, including a guidance law parameter and a first-order lateral time constant.
To this end, assuming that a missile with a proportional navigation (PN) guidance law attempts to attack an aerial target with bang-bang evasive maneuvers, a parameter identification model based on the gated recurrent unit (GRU) neural network is built in this paper.
The analytic identification solutions for the guidance law parameter and the first-order lateral time constant are derived.

The inputs of the identification model are available kinematic information between the aircraft and the missile, while the outputs contain the regression results of missile parameters.
To increase the training speed and the identification accuracy of the model, an output processing method called improved multiple-model mechanism (IMMM) is proposed in this paper.
The effectiveness of IMMM and the performance of the established model are demonstrated through numerical simulations under various engagement scenarios.
\end{abstract}

\begin{IEEEkeywords}
	Artificial Neural Network; Gated Recurrent Units; Identification; Multiple Model Mechanism
\end{IEEEkeywords}

\section{Introduction}
\label{section: Introducation}

\IEEEPARstart{I}{n} air combat, it is of great importance to protect high-value aircrafts from incoming missiles. Many passive defense methods have been proposed to increase the survivability of aircrafts, such as random telegraphing, and the periodic wave maneuver. \cite{AnAdaptiveTerminal1976, RootMeanSquare1996}
However, with the development of missiles and guidance laws, the performance of these strategies in modern air combat conditions is decreasing. In recent years, active defense methods have attracted considerable attention from researchers. \cite{CooperativeMultiple2010, CooperativeMissile2018, ActiveDefense2018, AdaptiveController2007}
Compared with aircrafts utilizing passive methods, aircrafts with active defense methods have higher survivability and better penetration capability.

Implementing an active defense method against a missile generally requires two key parameters of it: a guidance law parameter and a first-order lateral time constant.

The required parameters cannot be obtained by direct measurement.
% How to get N now
Therefore, several algorithms have been proposed to estimate and identify the aforementioned parameters.
In \cite{ActiveDefense2018}, an estimation approach based on the interactive multiple-model (IMM) method is proposed to identify the guidance law and states of a missile.
In \cite{AdaptiveController2007}, an adaptive receding horizon controller based on Bayesian inference is presented to identify the guidance law of a missile with perfect information on it.
% Most previous research only focus on N and use KF, and following drawbacks
Most previous studies only focused on identifying the guidance laws using extended Kalman filters (EKFs) \cite{CooperativeMultiple2010, Contribution1960, ANewApproach1960, NewResults1961, WingmanBased2020, AircraftGuidance2018} or unscented Kalman filters (UKFs) \cite{ActiveDefense2018, OptimalEstimation1988, NonlinearMMAE2018, MissileGuidance2014} under the assumption that the first-order lateral time constant is a known constant. However, there are several drawbacks of using a KF-based estimation method to solve this problem:

% it needs some information hard to get
a. It requires difficult-to-obtain information.
The KF is a predictor-corrector approach, whose calculation process can be generally divided into two steps: state prediction and state updating.
To predict the state of a system, some kinematic information about the missile, such as the velocity and velocity angle of the missile, must be accurately obtained. These information may not be accurate enough in complex and tense air combat situations.
The general countermeasure is to view the deviation between the true data and the data used for state prediction as a large process noise, which may lead to non convergence of the model.

b. To address the uncertainty of parameters (e.g., the speed of the missile, the drag coefficient, the guidance law and the first-order lateral constant), multiple-model adaptive estimators (MMAEs) \cite{OptimalAdaptive1965} and IMMs \cite{MultipleModel1999} are generally used in the developed estimation model.
Both methods are derived under the assumption that the system is in a known finite set of possible regimes, and the parameters of the true system are fixed (in MMAEs) or allowed to transition between regimes with a preset transition probability matrix (in IMMs).
If this assumption does not hold, additional models should be added to identify the true state of the system, which may increase the computational burden of the aircraft.

% constant velocity
c. The velocity of the studied missile is generally set as a constant to linearize the equation of motion and increase the speed of computation. The performance of the utilized identification model may deteriorate if the velocity of the missile changes considerably.

d. Due to drawback b and drawback c, to limit the size of the model and relieve the burden imposed on the computer, only the guidance laws  are generally set differently between regimes, while other parameters (such as the first-order lateral time constant and the constant speed) are considered known constants.

% introduction of ANN, LSTM, GRU
With the development of artificial intelligence (AI) technology, long short-term memory (LSTM) \cite{LongShort1997}, which is a kind of artificial neural network (ANN), has been widely used to solve time-related problems in aircrafts. \cite{MechineLearning2019, LearningTemporal2019, DecentralizedAutomotive2021, FastGuidance2021, MMM}.
Compared with MMAEs and IMMs, where each KF cooresponds to a specific scenario, the ANN is trained based on dataset, which contains samples extracted from various scenarios. Therefore, the generalization abiliey of ANN is higher than that of KFs when there are many parameters to be identified.
Moreover, the computing speed of ANN is faster since it doesn't require to calculate the process of every KF.
A simplified form of LSTM, which is called the gated recurrent unit, is proposed in \cite{LearningPhrase2014}. Compared with the conventional LSTM, the training speed of GRU is improved, while the accuracy is identical.
In \cite{FastGuidance2021}, an identification model based on a GRU neural network is established to identify the guidance law of an incoming missile. The model remains based on the assumption that the guidance law is in a finite set known by the aircraft. 
In \cite{MMM}, a regression guidance law identification model is built without adopting the above simplified assumption, but the model can only be used to indentify the guiance law parameter.

% The main contributation of this paper:
% 1.Tau first; 2.Regression, use MMM to limit the output in a reasonable range; 3. Identification effect under different N and Tau; 4. Influence of Cd
This paper is based on and is an improvement upon \cite{MMM}.
% Tau
A regression parameter identification model based on the GRU neural network is established in this paper. 
The inputs of the model are available information between the missile and the aircraft. In this case, the difficult-to-obtain information of the missile, which is required in KFs, is no long necessary.
The outputs of the model are regressoin results of both the guidance law parameter and the first-order lateral time constant. 
To the best of our knowledge, this is the first paper to identify the first-order lateral time constant in air combat. A theoretically feasible method is proposed in \cite{CooperativeMultiple2010}, but the method is difficult to implement in reality considering the aforementioned drawbacks of KFs.
% as previously mentioned
% b.MMM for multiple output
% advantage of IMMM
In addition, to increase the training speed and the identification accuracy of the established model, a neural network output processing method called the improved multiple-model mechanism is proposed, which can be applied to general multiple-output regression problems. 
Moreover, compared with a conventional neural network whose outputs may deviate too much from the required range and induce the system instability, the neural network using IMMM can ensure the outputs lie within a reasonable range, which increases the robustness of the system.

% structure of this paper
The remainder of this paper is organized as follows. The next section presents a nonlinear dynamic model of engagement and an analysis of the parameter identification problem. The concept of IMMM, structure of the parameter identification model and dataset establishment method are presented in \autoref{section: Identification Model}. A comprehensive performance analysis of the proposed IMMM and established identification model is presented in \autoref{section: Performance Analysis}, followed by concluding remarks.

% -------------------------------------------------------------------
% ------------------------Problem Formulation------------------------
% -------------------------------------------------------------------
\section{Problem Formulation}
\label{section: Problem Formulation}
In this section, we present a mathmatical model of the engagement, followed by a characteristics analysis of the parameter identification problem.

% ----------------------Nonlinear Dynamic Model----------------------
\subsection{Nonlinear Dynamic Model}
\label{subsection: Nonlinear Dynamic Model}
\begin{figure}[hbt!]
	\centering
	\includegraphics[width=.4\textwidth]{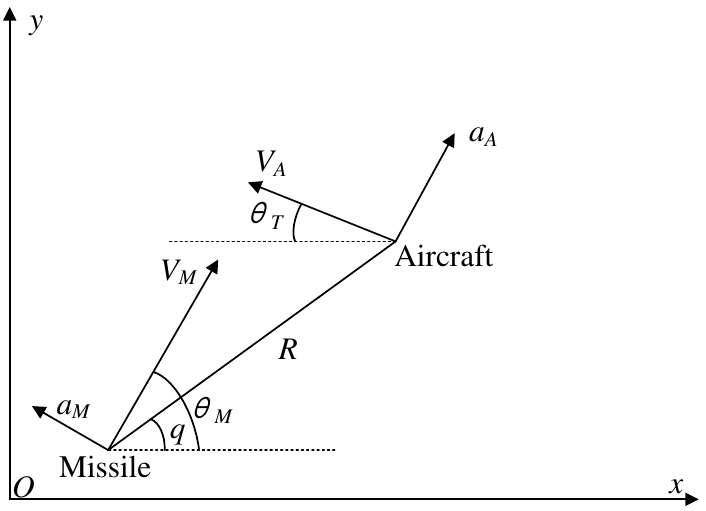}
	\caption{Cartesian inertial reference frame.}
	\label{fig: Cartesian inertial reference frame}
\end{figure}

% reference frame and denotation
We consider a scenario in which an incoming missile pursues an aircraft in a planar Cartesian inertial reference frame, as shown in \autoref{fig: Cartesian inertial reference frame}. Subscripts $ M $ and $ A $ denote variables associated with the missile and the aircraft, respectively. The velocity, normal acceleration, and flight-path angle are denoted by $ V $, $ a $ and $ \theta $, respectively. The missile-aircraft range and line-of-sight (LOS) angle are denoted by $ R $ and $ q $, respectively.

% relative kinematics
We assume that both vehicles are skid-to-turn roll-stabilized. Therefore, the engagement kinematics can be expressed in polar coordinates ($ R $, $ q $) as follows:

\begin{equation}
	\label{eq: Relative Equation}
	\begin{aligned}		
		V_R & = - [V_t \cos(\theta_A -q) + V_M \cos(\theta_M - q)] \\
		\dot{q} & = \frac{V_t \sin(\theta_A - q) - V_M \cos(\theta_M - q)} {R}
	\end{aligned}	
\end{equation}

\noindent where $ g = 9.8m/s^2 $ is the gravity coefficient.

% first-order lateral
We also assume the first-order lateral maneuver dynamics for both missile and aircraft, i.e.,

\begin{equation}
	\label{eq: first-order lateral}
	\dot{a_i} = \frac{a_{c,i} - a_i}{\tau_i}, i \in {A, M}
\end{equation}

\noindent where $a_c$ is the acceleration command and $\tau$ is the first-order lateral time constant.

% PN guidance law
The missile is assumed to employ the PN guidance law. The acceleration command $ a_c $ is

\begin{equation}
	\label{eq: PN}
	a_c = N \dot{q} V_R
\end{equation}

\noindent where $ N $ is the PN guidance law parameter and $V_R$ is the relative velocity between the missile and the aircraft.

% drag
Additionally, the effect of the external force on the velocity of the missile is considered. The dynamics can be expressed as:
\begin{align}
	\label{eq: dotV}
	\dot{V} &= (T - D) / m - g \sin \theta_M \\
	\label{eq: D}
	D &= (\rho V^2 / 2) C_D S_M
\end{align}
\noindent where $T$ is the thrust force; $D$ is the drag force; $m$ is the mass of missile; $\rho$ is the density of air; $C_D$ is the drag coefficient; and $S_M$ is the reference area.

\subsection{Measurement Model}
\label{subsection: Measurement Model}

% aircraft own state
We assume that the inertial vector of the aircraft
\begin{equation}
	\label{eq: x_A}
	x_A = [\theta_A, a_A, V_A] ^ T
\end{equation}
\noindent can be known with high accuracy via a navigation system.\cite{CooperativeMultiple2010}

% measurement
The aircraft is assumed to be equipped with a radar seeker. The direct measurement data include the relative distance $R$ and LOS angle $q$. We also assume that the measurements $M_t$, $t \in N $ are mutually independent and can be acquired at discrete time instances $tT_p$, where $T_p > 0$ is a fixed measurement period. Meanwhile, $M_t$ are assumed to be contaminated by zero-mean white Gaussian noise. Thus, the measurement model can be expressed as:
\begin{equation}
	\label{eq: Measurement and Noise}
	M_t = \begin{bmatrix} R(t) \\ q(t)  \end{bmatrix} + \nu(t)
\end{equation}
\noindent where
\begin{equation}
	\nu(t) \sim \mathcal{N}([0]_{2\times1}, Q),
	Q = diag(\sigma_R^2, \sigma_q^2)
\end{equation}

% -------------Analysis of Identification Problem--------------------
\subsection{Analysis of the Parameter Identification Problem}
\label{subsection: Analysis of the Identification Problem}
% analysis and compare the difference of the direct calculation, Kalman Filter and ANN

\begin{figure}[hbt!]
	\centering
	\includegraphics[width=.45\textwidth]{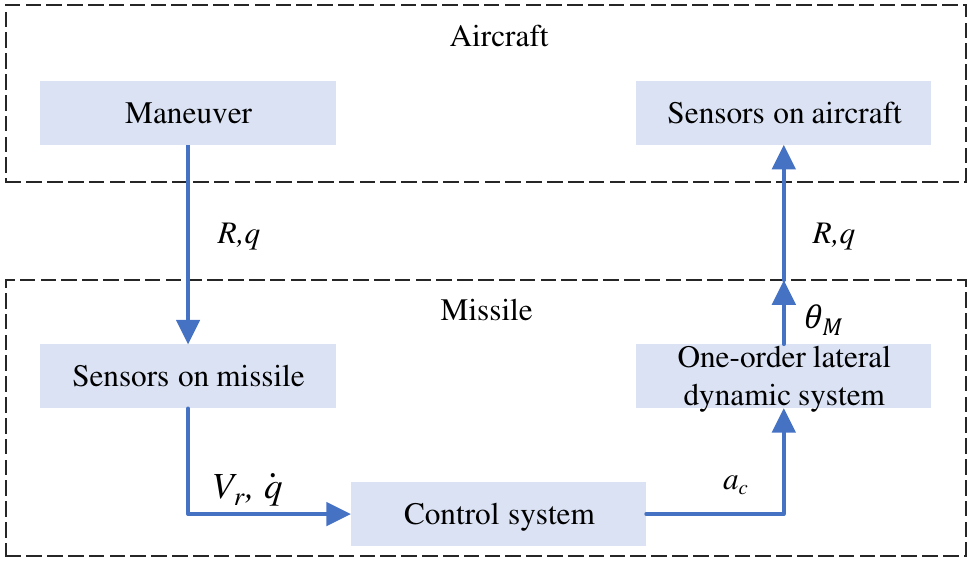}
	\caption{Diagram of the data flow.}
	\label{fig: data flow}
\end{figure}

The data flow of the engagement is shown in \autoref{fig: data flow}. The control system on the missile outputs an acceleration command according to the data from onboard sensors. 

% calculate theta_M and V_M
In \eqref{eq: Relative Equation}, the directly accessible information about the aircraft include $V_t$, $\theta_A$, $R$, $q$, $\dot{q}$. Therefore, the equations can be written as:
\begin{gather}
	\label{eq: V_M cos}
	V_M \cos(\theta_M - q) = f_1(V_R, V_A, \theta_A, q)  \\
	\label{eq: V_M sin}
	V_M \sin(\theta_M - q) = f_2(\dot{q}, R, V_A, \theta_A, q)
\end{gather}
\noindent where
\begin{gather}
	\label{eq: f1}
	f_1(V_R, V_A, \theta_A, q) = -V_R - V_A \cos(\theta_A + q) \\
	\label{eq: f2}
	f_2(\dot{q}, R, V_A, \theta_A, q) = - \dot{q} R + V_A \sin(\theta_A + q)
\end{gather}

Dividing \eqref{eq: V_M sin} by \eqref{eq: V_M cos} yields
\begin{equation}
	\label{eq: theta_M}
	\theta_M = \arctan \frac{f_2(\dot{q}, R, V_A, \theta_A, q)}{f_1(V_R, V_A, \theta_A, q)} + q
\end{equation}

Substituting \eqref{eq: theta_M} into \eqref{eq: V_M cos} yields
\begin{equation}
	\label{eq: V_M}
	V_M = f_1(V_R, V_A, \theta_A, q) \cos(\theta_M - q) + f_2(\dot{q}, R, V_A, \theta_A, q) \sin(\theta_M - q)
\end{equation}

% acceleration of missile
The acceleration of the missile can be calculated according to data obtained from \eqref{eq: theta_M} and \eqref{eq: V_M}:
\begin{equation}
	\label{eq: a}
	a_M = \frac{V_M}{g} \dot{\theta}_M +  \cos \theta_M  
\end{equation}

Substituting \eqref{eq: PN} into \eqref{eq: first-order lateral} yields:
\begin{equation}
	\label{eq: tau_M}
	\tau_M = \frac{N V_{R,t} \dot{q}_t - a_t}{a_t - a_{t-1}} 
		= \frac{N V_{R,t-1} \dot{q}_{t-1} - a_{t-1}}{a_{t-1} - a_{t-2}} =...
		= \frac{N V_{R,1} \dot{q}_{1} - a_{1}}{a_{1} - a_{0}}
\end{equation}
\noindent where subscripts $t$, $t-1$, $t-2$, ..., $1$, $0$ are discrete time instances.

% noise of the question
\begin{remark}
	The analytic solutions to $N$ and $\tau_M$ can be directly calculated using above equations. However, it is hard to guarantee the accuracy of the required data considering the measurement noise. Therefore, the direct calculation of $N$ and $\tau_M$ may not work.
\end{remark}

According to \eqref{eq: theta_M} to \eqref{eq: tau_M}, the identification question can be written as:
\begin{equation}
	\label{eq: f, queation}
	[\tau_M, N] = f(R, q, \theta_A, V_a, \nu)|_{(0, t)}
\end{equation}
\noindent where $f$ represents a complex nonlinear mapping relationship and subscript $(0, t)$ represents that the question is related to time. Thus, the key to solving the question turns into finding the mapping relationships between the kinematic information and parameters.

% -------------------------------------------------------------------
% -----------------------Identification Model------------------------
% -------------------------------------------------------------------
\section{Identification Model}
\label{section: Identification Model}
In this section, we firstly present the concept of IMMM, which can be applied to general multiple-output regression problems. Then, we present the structure of the parameter identification model, followed by dataset establishment.

\begin{notation}
	For simplicity, in the remainder of this paper, we denote the $sigmoid$, $tanh$, and $softmax$ functions in italics, as follows:
	\begin{gather*}
		sigmoid(\zeta) = \frac{1}{1+e^{- \zeta}}  \\
		tanh(\zeta) = \frac{e^{\zeta} - e^{- \zeta}}{e^{- \zeta}+e^{\zeta}}  \\
		softmax(\zeta_i) = \frac{e^{\zeta_i}}{\sum_j e^{\zeta_j}}
	\end{gather*}
\end{notation}

\subsection{Improved Multiple-Model Mechanism}
\label{subsection: Multiple-Model Mechanism}
The multiple-model mechanism was proposed by Wang et al. in \cite{MMM}. 
The main idea of MMM is that use regimes to represent different possible situations by connecting a multiple-model layer behind a conventional neural network.
However, the conventional MMM can only be used in single-output problems. To solve this drawback, we introduce the thought of transfer learning to MMM and propose an output processing method called the improved multiple-model mechanism. 

The main idea of the IMMM is to set several groups of regimes in the multiple-model layer, with each group corresponding to a regression result of the neural network. Regimes in different groups represent possible situations of different regression results. A certain regresssion result is the weighted sum of regimes in the corresponding group.

\begin{figure*}[hbt!]
	\centering
	\includegraphics[width=0.7\textwidth]{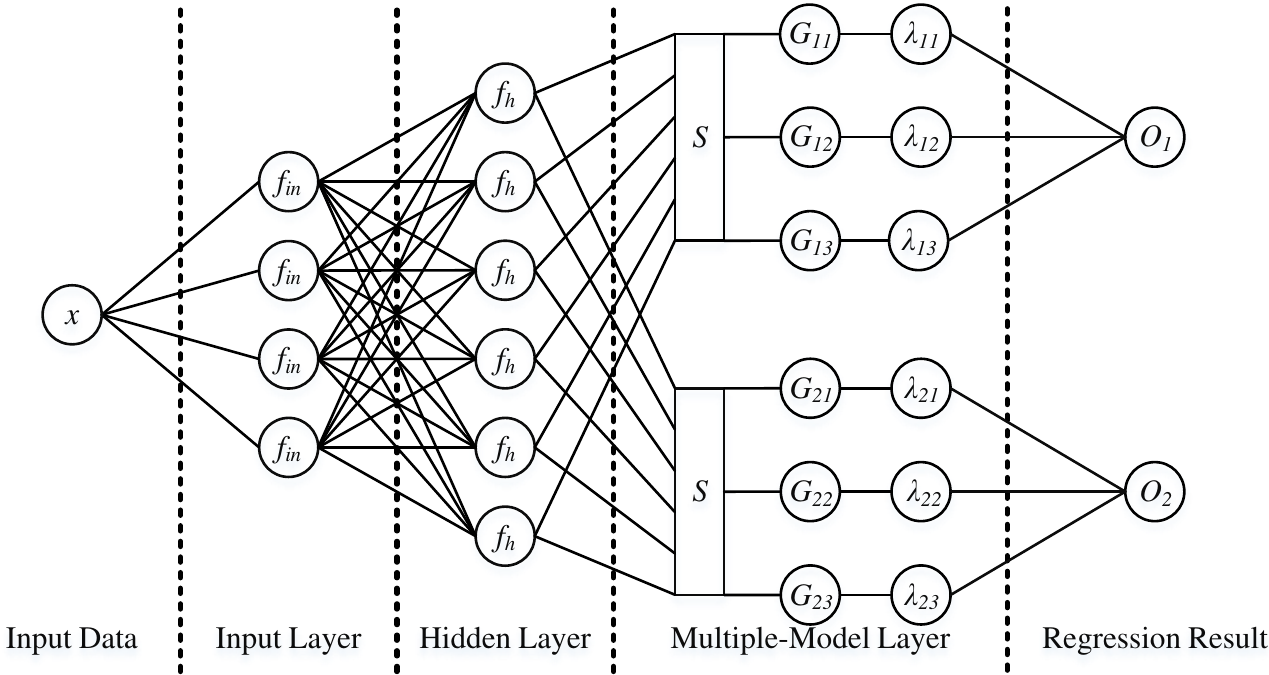}
	\caption{Improved multiple-model mechanism in the multiple-output problem.}
	\label{fig: IMMM}
\end{figure*}

The structure of a neural network using the IMMM is shown in \autoref{fig: IMMM}. There are three types of layers: an input layer, several hidden layers and a multiple-model layer. The input layer is used to process the input data onto an appropriate range. General activation functions of the input layer include $sigmoid$ and $tanh$, whose output ranges are $[0.0, 1.0]$ and $[-1.0, 1.0]$, respectively. Hidden layers are the main part of the neural network, which are composed of different kinds of neurons to extract features from samples. The multiple-model layer is the main difference between a conventional neural network and a neural network using the IMMM. Regimes representing different situations are set in the multiple-model layer, and weights of different regimes are calculated through the $softmax$ function as follows:
\begin{equation}
	\label{eq: multiple-model layer}
	G_i = [G_{i,1}, G_{i, 2}, G_{i, p_i}]^T = softmax(w_{out}h_{last} + b_{out}),
\sum_{j=1}^{p_i} G_{i,j} = 1
\end{equation}
\noindent where $G_{ij}$ is the $j$th regime's weight of the $i$th regression result;  $p_i$ is the number of regime of the $i$th regression result; $w_{out}$ is the weight matrix between the last hidden layer and the multiple-model layer; $h_{last}$ is the output of the last hidden layer; and $b_{out}$ is the bias vector. There is more than one group of regimes. Each group of regimes corresponds to a regression result, and the sum of a group of regimes is 1.

%\begin{multicols}{2}
%\end{multicols}

\begin{remark}
	The connection between the last hidden layer and the multiple-model layer is separated into several groups instead of a full connection.
\end{remark}

Regression results of the neural network using IMMM are the weighted sum of the corresponding group of regimes:
\begin{equation}
	\label{eq: multiple-model layer sum}
	O_i = \Lambda _i^T G_i = \sum_{j=1}^{p_i} \lambda _{i,j} G_{i,j}
\end{equation}
\noindent where $O_i$ is the $i$th regression result and $\lambda_{i,j}$ is the $j$th regime of the $i$th regression result. 

Compared with those of a conventional neural network, the training speed and the accuracy of a neural network using IMMM is increased because the initial output ranges of the latter are limited in a required range before training, as shown in \eqref{eq: multiple-model layer sum}.
Moveover, the useage of IMMM can ensure the outputs of the neural network lie within a reasonbale range, which increases the robustness of the system.

\begin{remark}
	The principle of transfer learning shows that the front layers of a neural network are generally used to extract primary features, while the posterior layers are used to deeply analyze samples. Therefore, the model in one problem can be used in other similar problems by merely training its last layer. The IMMM is based on this idea. Only the output of the last hidden layer is used to calculate the weights of regimes.
\end{remark}

\subsection{Structure of the Parameter Identification Model}
\begin{figure}[hbt!]
	\centering
	\includegraphics[width=.45\textwidth]{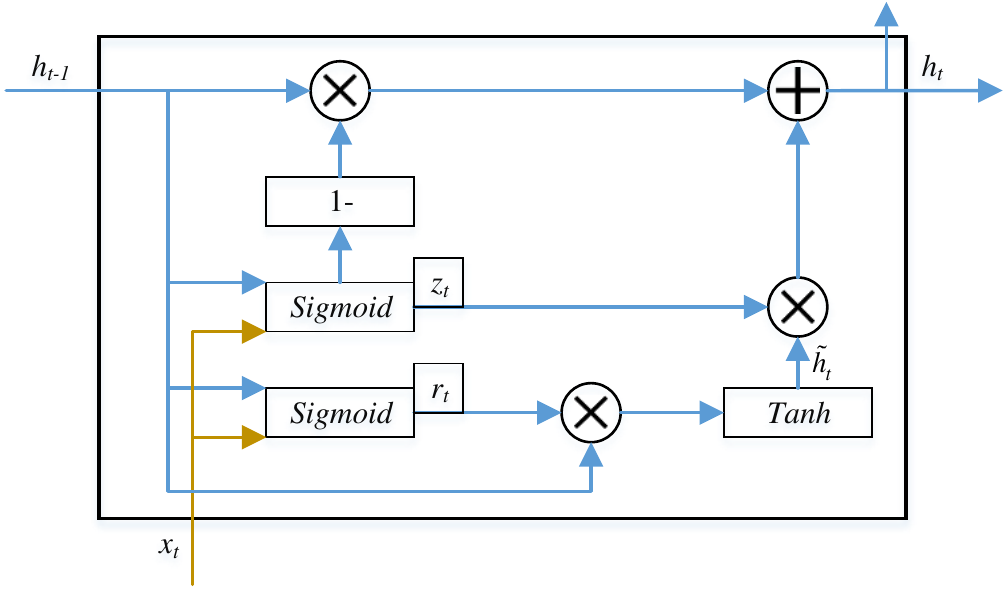}
	\caption{Basic GRU neuron.}
	\label{fig: BasicGRU}
\end{figure}

% basic concept of gru
The parameter identification model established in this paper is based on the GRU neural network using IMMM .
The diagram of a basic GRU neuron is shown in \autoref{fig: BasicGRU}, which is composed of four parts: a reset gate ($r_t$), an updata gate ($z_t$), a candidate state ($\widetilde{h}_t$) and an output state ($h_t$).

Calculation equations of a GRU neuron are as follows:
\begin{gather}
	\label{eq: r_t}
	r_t = sigmoid(w_{hr} \bullet h_{t-1} + w_{xr} \bullet x_t + b_r)  \\
	\label{eq: z_t}
	z_t = sigmoid(w_{hz} \bullet h_{t-1} + w_{xz} \bullet x_t + b_z)  \\
	\label{eq: hat h_t}
	\widetilde{h}_t = tanh(w_{h\widetilde{h}_t} \bullet r_t \times h_{t-1} + w_{x\widetilde{h}} \bullet x_t + b_{\widetilde{h}})  \\
	\label{eq: h_t}
	h_t = (1 - z_t) \times h_{t-1} + z_t \times \widetilde{h}_t
\end{gather}
\noindent where $w$ is the weight matrix; $b$ is the bias vector; $\bullet$ denotes a matrix multiplication; and $\times$ denotes an element-wise multiplication.
As can be seen from \eqref{eq: r_t} to \eqref{eq: h_t}, the value of $z_t$ signifies the reliance on the previous information. A larger value corresponds to a smaller weight of the newly-inputting information.

\begin{figure*}[htb!]
	\centering
	\includegraphics[width=0.99\textwidth]{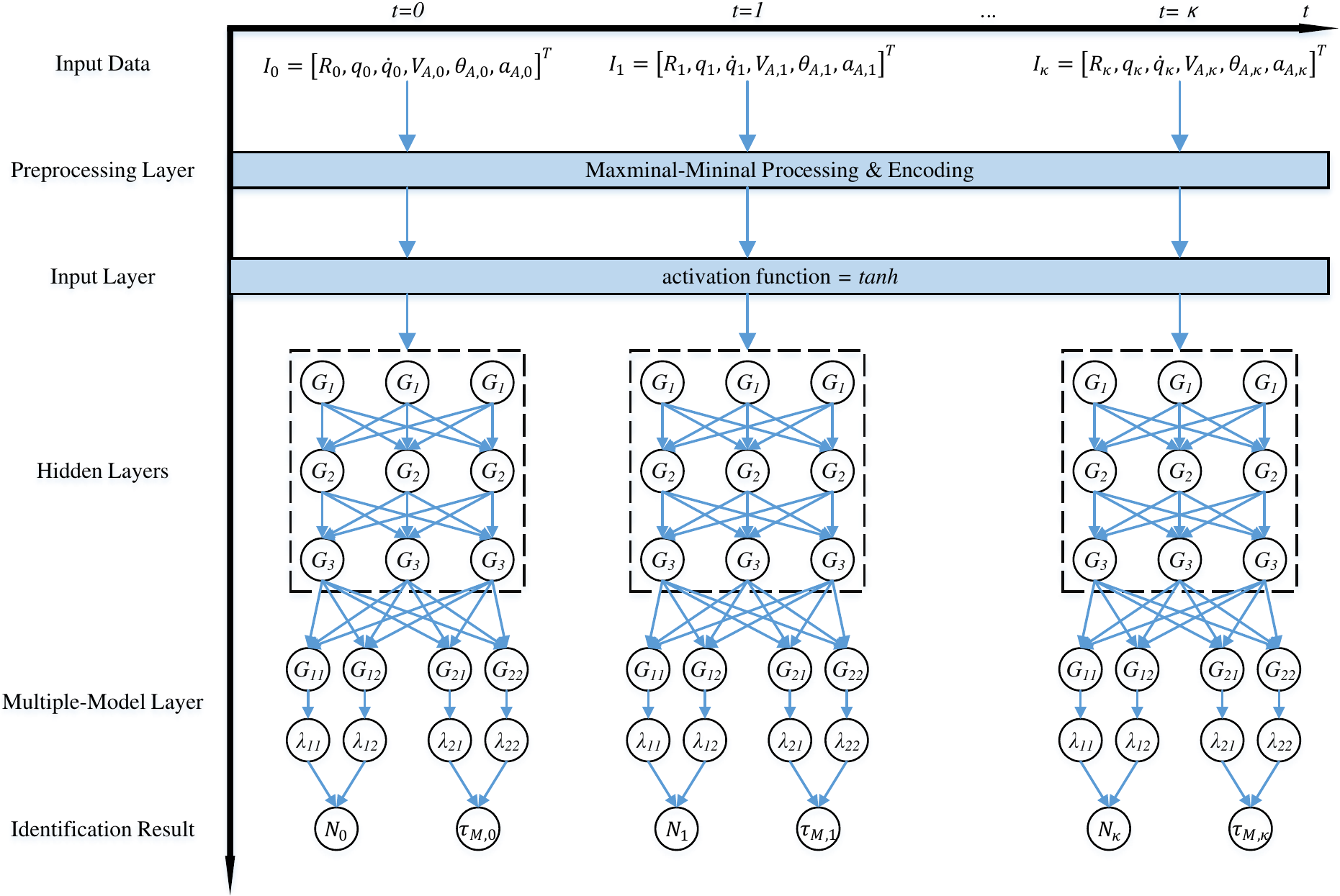}
	\caption{Structure of the identification model.}
	\label{fig: Structure of model}
\end{figure*}

\label{subsection: Structure of Identification Model}
The structure of the parameter identification model based on the GRU neural network using IMMM is shown in \autoref{fig: Structure of model}, 
where each subscript $t \in N$ represents a specific time; 
%where each subscript $t \in N$ represents the information at the specific time; 
$I_t$ is inputs of the model; $G_{i,j}$ is the $j$th regime's weight of the $i$th parameter, where $i=1,2$ denote the guidance law parameter $N$ and the first-order lateral time constant $\tau_M$, respectively; and $\lambda_{i,j}$ is the $j$th regime of the $i$th parameter.

Inputs of the parameter identification are available kinematic information between the aircraft and the missile, including
\begin{equation}
	\label{eq: input data}
	I_t = [R_t, q_t, \dot{q}_t, V_{A,t}, \theta_{A,t}, a_{A,t}]^T
\end{equation}

The preprocessing layer is used to process and encode the input data.
The input step $l$ is set according to the preset input step of model $K$ and the length of the input period $\kappa$:
\begin{equation}
	\label{eq: encode}
	l = \min(K, \kappa)
\end{equation}
\begin{equation}
	\label{eq: encode 2}
	x_1, x_2, ..., x_l = \left\{
	\begin{array}{lr}
	I_1, I_2, ..., I_{\kappa}, \kappa < K  \\
	I_{\kappa-K+1}, I_{\kappa-K+2}, ..., I_{\kappa}, \kappa \ge K
	\end{array}
	\right.	
\end{equation}

The input data is normalized into dimensionless values with the same range using min-max normalization:
\begin{equation}
	\label{eq: man-min}
	\zeta(i) = \frac{\zeta'(i) - I(i)_{min}}{I(i)_{max} - I(i)_{min}}
\end{equation}
\noindent where $\zeta(i)$ is the processed data; 
$\zeta'(i)$ is the initial data; 
%$I(i)$ is the $i$th input data of the initial input data vector; 
$I(i)_{min}$ is the minimum value of the $i$th input data; and $I(i)_{max}$ is the maximum value of the $i$th input data.

The activation function of the input layer is $tanh$, which is used to compress or expand the input range to $[-1,1]$ and increase the nonlinear characteristic of the model.
Neurons in hidden layers are basic GRU neurons, whose structure and calculation equations are shown in \autoref{fig: BasicGRU} and in \eqref{eq: r_t} to \eqref{eq: h_t}, respectively. The multiple-model layer is built according to the structure in \autoref{fig: IMMM}. The weights of different regimes are calculated by \eqref{eq: multiple-model layer sum}. 
The maximum and minimum regimes of each group are set as the maximum and minimum values of corresponding data in the training dataset respectively.
Outputs of the model are identification results of the guidance law parameter $N$ and the first-order lateral time constant $\tau_M$.

\subsection{Dataset Establishment}
\label{subsection: Dataset Establishment}
The dataset is built based on the nonlinear dynamic model in \autoref{subsection: Nonlinear Dynamic Model} using Latin hypercube sampling.
Compared with the Monte Carlo simulation, Latin hyper sampling can increase the generalization ability of the dataset. Samples are extracted from engagement simulations using slipping windows. Inputs of the samples are shown in \eqref{eq: input data}, while labels are identical to the outputs of the model. In addition, the labels are processed into the same range using max-min normalization.

\begin{remark}
	The Latin hypercube sampling is used twice in the process of building the dataset. It is firstly used to set the initial parameters of engagements and subsequently used to extract samples from the engagement simulations.
\end{remark}

% -------------------------------------------------------------------
% -----------------------Performance Analysis------------------------
% -------------------------------------------------------------------
\section{Performance Analysis}
\label{section: Performance Analysis}
In this section, the effectiveness of IMMM and the performance of the established parameter identification model are verified by numerical simulations. We firstly present simulation parameters and engagement scenarios.
Secondly, the training comparison between a conventional identification model and an identification model with IMMM is present
Then, we present an sample identification run and several Monte Carlo simulations to demonstrate the performance of the established identification model. The identification accuracies under different drag coefficients are presented at the end of this section.

% ----------------Simulation Parameters and scenario----------------------
\subsection{Simulation Parameters and Scenarios}
\label{subsection: Simulation Environment and Scenario}
\begin{remark}
	\label{remark: Simulation parameters same}
	For both dataset establishment and following simulations, the simulation parameters and engagement scenarios are set in identical manners as follows.
\end{remark}

% initial position
The initial distance and LOS angle between the missile and the aircraft are $R(0) \in [6000, 8000]$ and $q(0) \in [0, 5] \deg$, respectively.
% aircraft
The aircraft performs a bang-bang maneuver during the engagement, with a first-order lateral time constant $\tau_A = 0.6s$, an initial velocity angle $\theta_A = 0\deg$,  and a constant velocity $V_A \in [0.8, 1.0] Ma$, where $Ma = 340m/s$ is the velocity of sound. The amplitude and the frequency of the bang-bang maneuver are $\eta = 8g$ and $\xi = 1/8$, respectively.

\begin{figure}[hbt!]
	\centering
	\includegraphics[width=.45\textwidth]{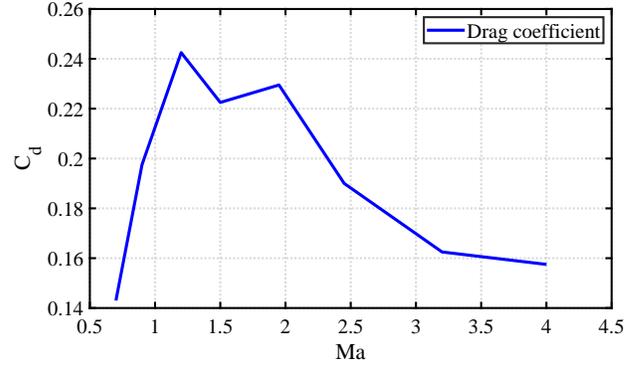}
	\caption{Drag coefficient.}
	\label{fig: Drag Coefficient}
\end{figure}

% missile
The missile is launched at $t = 0$, with an initial velocity $V_M(0) \in [2.0, 2.5] Ma$, an initial velocity angle $\theta_M(0) = 0\deg$, and a reference area $S = 0.101 m^2$. The drag coefficient of the missile is shown in \autoref{fig: Drag Coefficient}. 
The guidance law parameter and the first-order lateral time constant of the missile are $N \in [2.5, 5.5]$ and $\tau_M \in [0.1, 0.4]$, respectively.

% measurement
The measurement sampling rate of the aircraft's radar is $f_m = 100Hz$. The measurement noises are $\sigma_R = 5m$ and $\sigma_q = 1mrad$.
The simulations are performed on TensorFlow-1.13.0, and the graphic card is GeForce RTX 3090. The batch size and the number of training iterations are set as 3000 and 100000, respectively. The initial learning rate is 0.002. The dacay rate of the learning rate is 0.99/100.

% --------------------Training Process of the Model-----------------------
\subsection{Performance of IMMM}
\label{subsection: Performance of IMMM}
\begin{remark}
	\label{remark: Train without noise}
	The inputs of the samples used for training are not contaminated by noise.
\end{remark}

The training comparison between a conventional identification model and an identification model with IMMM is present in this subsection.  The structure of the former model is built based on the conventional model presented in [CJA]. The number of hidden layers is 3, with each hidden layer containing 96 basic GRU neurons. The preset input time length is 1s. The loss function of both models is mean square error (MSE). Models in the multiple-model layer are
\begin{gather}
	\Lambda_1 = [2.50, 3.25, 4.00, 4.75, 5.50]^T \\
	\Lambda_2 = [0.100, 0.175, 0.250, 0.325, 0.400]^T
\end{gather}

\begin{figure}[hbt!]
	\centering
	 \subfigure[The beginning of the training process]{\includegraphics[width=.38\textwidth]{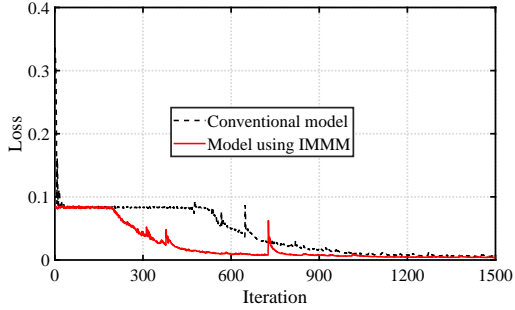}}
	 \subfigure[The end of the training process]{\includegraphics[width=.38\textwidth]{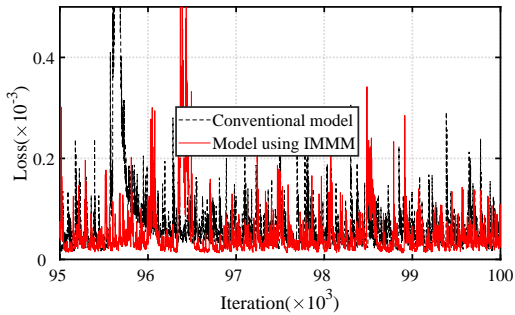}}
%	 \subfigure[0-100000]{\includegraphics[width=.4\textwidth]{NumModel_End}}
	\caption{Training comparsion between the conventional model and the model using IMMM}
	\label{fig: num of model}
\end{figure}

The training process is present in \autoref{fig: num of model} and \autoref{tab: num of model}. Note that compared with that of the conventional identification model, the initial MSE of the model using IMMM reduces from 0.3366 to 0.0834. This can be explained by the fact that the initial output ranges of the model using IMMM are limited in the required range before training, as shown in \eqref{eq: multiple-model layer sum}.
It can also be found that the MSE after training of the model using IMMM is lower than that of the conventional identification model, which means that the former model has a higher accuracy. This phenomenon validates the theoretical analysis in \autoref{subsection: Multiple-Model Mechanism}.

\begin{table*}[hbt!]
	\caption{MSE of the conventional model and the model using IMMM.}
	\centering
	\renewcommand\arraystretch{1.0}
	\begin{tabular}{ccc}
		\hline
		Model & Initial MSE & MSE after training ($\times 10^{-5}$)  \\
		\hline
		Conventional Model & 0.3366  & 10.5 \\
		Model using IMMM & 0.0834  & 5.16 \\
		\hline
	\end{tabular}
	\label{tab: num of model}
\end{table*}

\subsection{Sample Run}
\label{subsection: Sample Run}
% sample
To intuitively demonstrate the performance of the established parameter identification model, a sample example run is presented in this subsection before turning to statistical Monte Carlo simulations.
The initial distance and initial LOS angle for the sample run are $R(0) = 7000m$ and $q(0) = 0\deg$, respectively. The constant velocity of the aircraft is $V_A = 0.9Ma$. The guidance law parameter and the first-order lateral time constant of the missile are $N = 5.0$ and $\tau_M = 0.30$, respectively.

\begin{figure*}[t]
	\centering
	\subfigure[Weights of PN regimes]
	{\includegraphics[width=.45\textwidth]{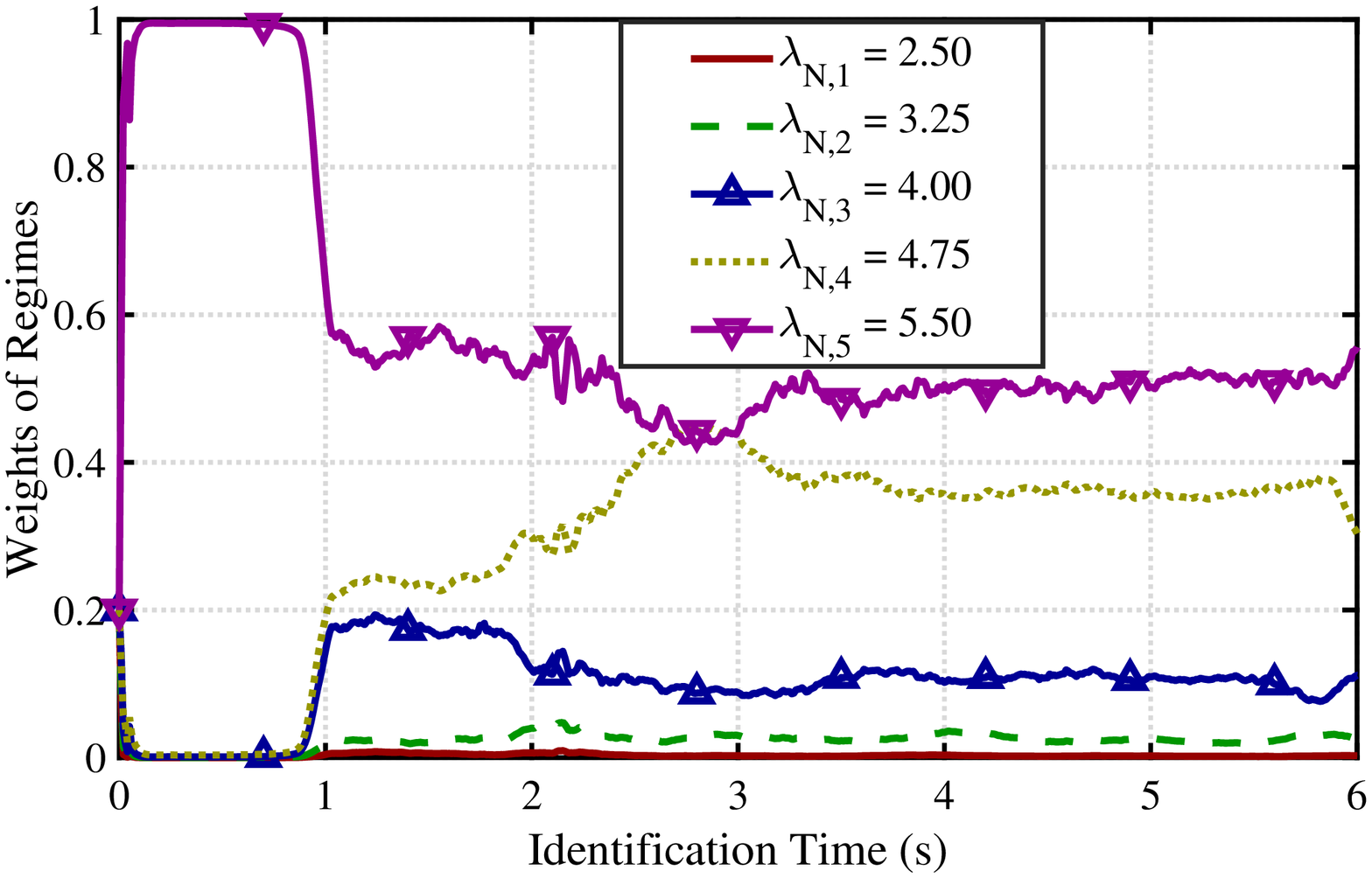}}
	\subfigure[Identification result of the PN parameter]
	{\includegraphics[width=.45\textwidth]{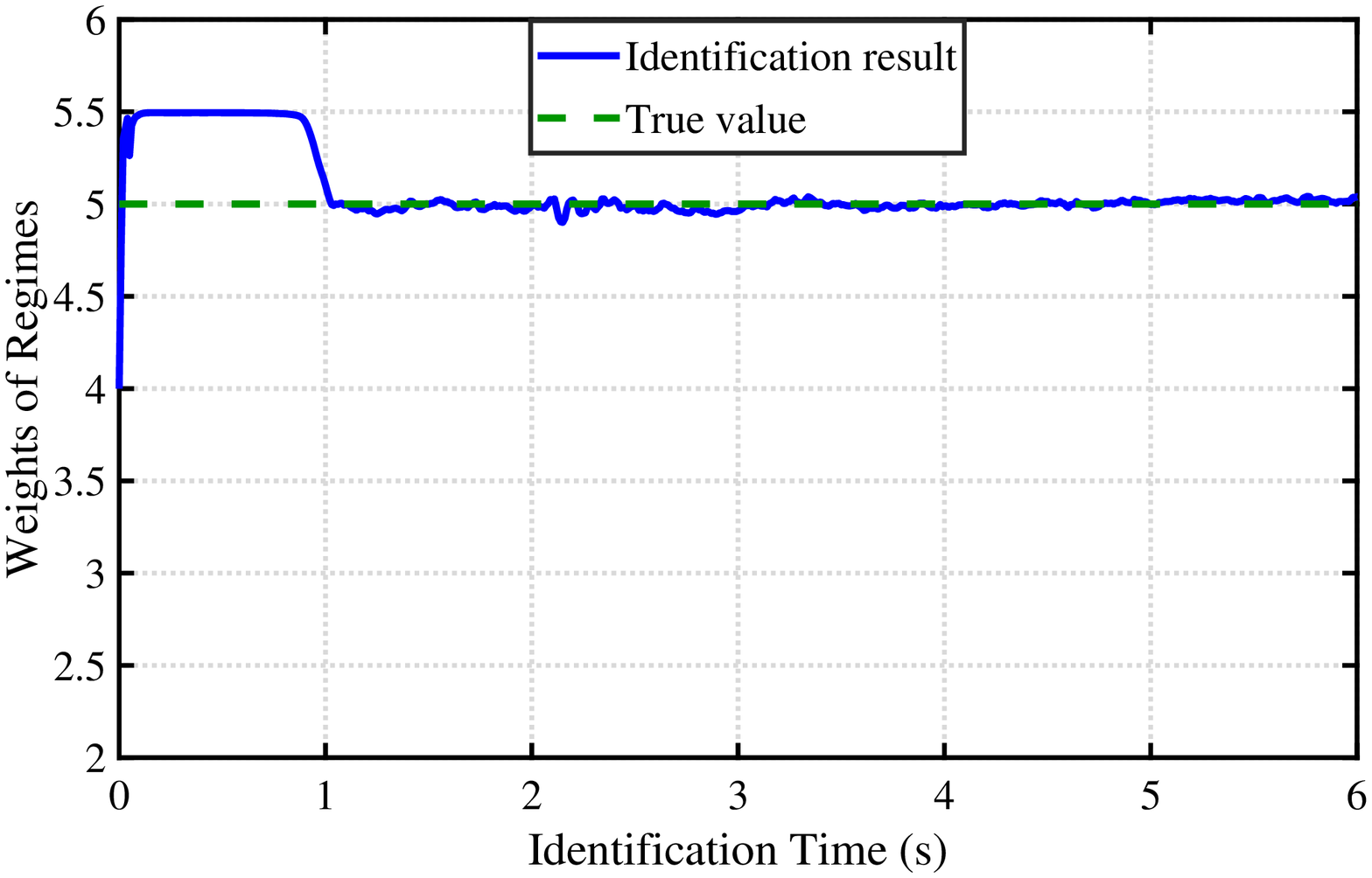}}
	
	\subfigure[Weights of $\tau_M$ regimes]
	{\includegraphics[width=.45\textwidth]{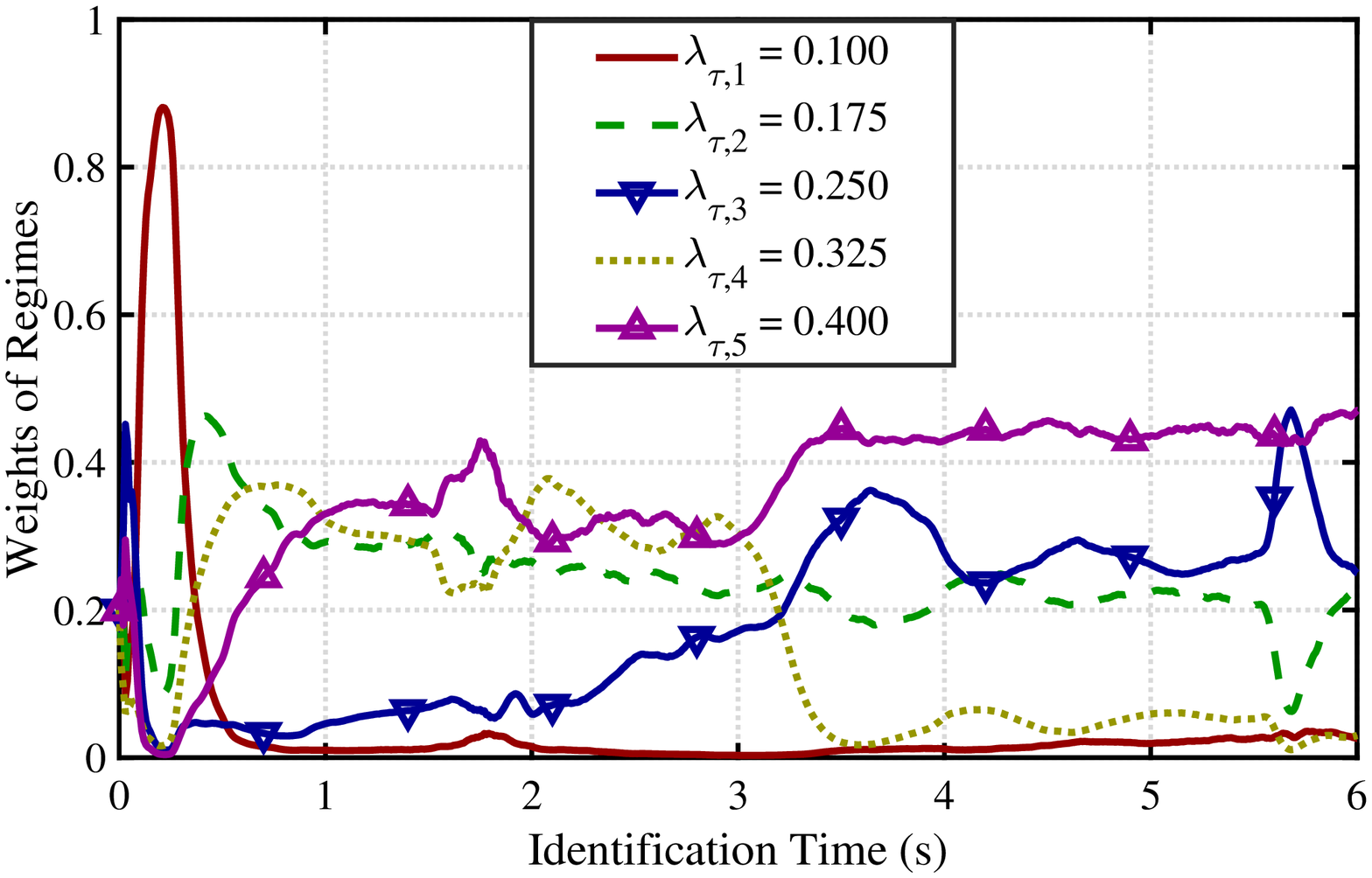}}
	\subfigure[Identification result of $\tau_M$]
	{\includegraphics[width=.45\textwidth]{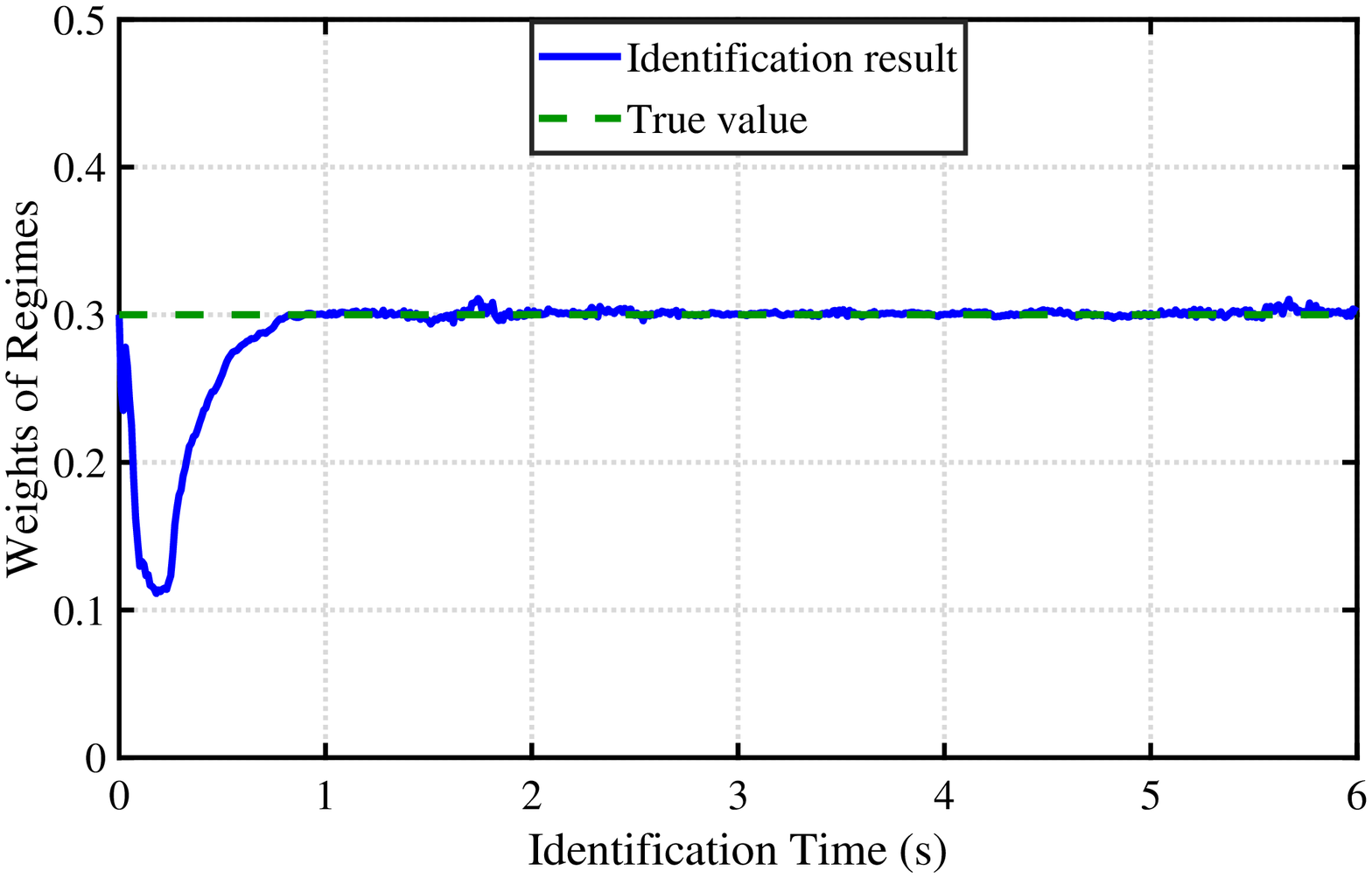}}
	\caption{Transition process of the parameter identification model during the sample run.}
	\label{fig: Transition of the identification model}
\end{figure*}

The transition process of weights of different regimes and the identification results of the model are presented in \autoref{fig: Transition of the identification model}. The weights of different regimes are identical at the beginning of the simulation; thus, the initial identification results for $N$ and $\tau_M$ are 4.0 and 0.25, respectively. 
It can be found that identification results of both the guidance law parameter and the first-order lateral time constant converge after 1 second. Then the outpus of the model fluctuate near the true value, which verifies the accuracy and the stability of the parameter identification model.

Note that in IMMM, the weights of regimes change constantly during the simulation, while the weighted sum of them, i.e., the identification results, remain stable. This is different from IMM or MMAE, in which the weight of the true situation converges to nearly $100\%$ at the end of the simualation\cite{CooperativeMultiple2010, AircraftGuidance2018}.

\subsection{Monte Carlo Simulation}
\label{subsection: Monte Carlo Simulation}

\begin{figure}[hbt!]
	\centering
	\includegraphics[width=0.4\textwidth]{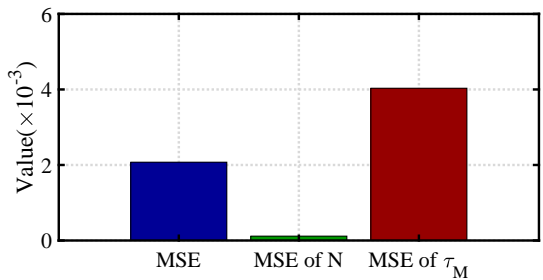}
	\caption{Results of the Monte Carlo simulation.}
	\label{fig: Monte Carlo simulation}
\end{figure}

A Monte Carlo simulation with 6000 independent runs is performed to show the performance of the established identification model. In each simulation, the initial parameters are randomly set from the corresponding ranges in \autoref{subsection: Simulation Environment and Scenario}. The samples used for the Monte Carlo simulation are extracted from trajectories that are not included in the training dataset, which can show the robustness and the generalization ability of the model.
The results of the simulation are presented in \autoref{fig: Monte Carlo simulation}.
The MSE of $N$ is $0.1158 \times 10^{-3}$, while the MSE of $\tau_M$ is $4.0327 \times 10 ^{-3}$. This implies that the model has a better performance in identifying the guidance law parameter.

An extra Monte Carlo simulation with different $N$ and $\tau_M$ is carried on to explore the performance of the model under different situations. The guidance law parameters are
\begin{equation*}
	N \in \{2.5, 2.8, 3.1, 3.4, 3.7, 4.0, 4.3, 4.6, 4.9, 5.2, 5.5\}
\end{equation*}
\noindent and the first-order lateral time constants are
\begin{equation*}
	\tau_M \in \{0.10, 0.13, 0.16, 0.19, 0.22, 0.25, 0.28, 0.31, 0.34, 0.37, 0.40\}
\end{equation*}

The other parameters are randomly set according to \autoref{subsection: Simulation Environment and Scenario}. The number of situation is $11 \times 11 = 121$, and the number of trajectories of each situation is 150. 
Randomly extracted 20 samples from each trajectory. 
Thus the total number of samples used for this Monte Carlo simulation is 363,000.
The performances of the parameter identification model is presented in Tables \ref{tab: N and Tau with noise}.
As the tables show, the model has good performance under all conditions, which verifies the accuracy and the robustness of the parameter identification model.

\begin{table*}
	\centering
	\renewcommand\arraystretch{1.0}	
	\caption{Performance of the model under different conditions}	
	\begin{tabular}{c|ccccccccccc}
		\hline
		\diagbox{$N$}{MSE ($10^{-3}$)}{$\tau_M$} & 0.10  & 0.13  & 0.16  & 0.19  & 0.22  & 0.25  & 0.28  & 0.31  & 0.34  & 0.37  & 0.40 \\
		\hline
	    2.5   & 2.1049 & 2.1683 & 2.1378 & 2.1083 & 2.1668 & 2.0301 & 2.0728 & 1.8673 & 2.1577 & 2.0635 & 1.9578 \\
		2.8   & 2.0720 & 2.0066 & 2.0416 & 2.0121 & 2.1872 & 2.1645 & 2.2958 & 2.0286 & 2.1795 & 2.0532 & 2.1917 \\
		3.1   & 2.1323 & 2.3942 & 2.0697 & 1.8303 & 2.0843 & 2.0447 & 2.1308 & 1.9942 & 2.0125 & 2.1073 & 1.9877 \\
		3.4   & 2.0605 & 1.9035 & 1.7985 & 2.1465 & 1.9790 & 2.1042 & 2.1197 & 2.0228 & 2.0160 & 1.9873 & 2.1099 \\
		3.7   & 2.0874 & 2.2401 & 2.0292 & 2.0027 & 2.2811 & 2.1474 & 2.1627 & 1.9145 & 2.1496 & 2.0768 & 1.9666 \\
		4.0   & 1.9881 & 1.9516 & 1.9889 & 2.2014 & 2.0575 & 1.9308 & 2.1153 & 1.9723 & 1.9161 & 2.0797 & 2.1350 \\
		4.3   & 1.9425 & 2.0073 & 1.9604 & 2.0295 & 1.9342 & 2.1026 & 1.8952 & 2.0608 & 2.1089 & 1.8242 & 2.0253 \\
		4.6   & 2.0207 & 2.0896 & 2.1126 & 2.0661 & 2.2251 & 1.8649 & 2.1107 & 1.8768 & 2.0950 & 1.8525 & 1.8694 \\
		4.9   & 2.1794 & 2.0254 & 2.0687 & 2.1032 & 1.8923 & 2.0192 & 2.1390 & 2.0725 & 2.1583 & 2.0566 & 2.0050 \\
		5.2   & 1.9426 & 1.9433 & 1.7845 & 2.1938 & 2.0058 & 2.0076 & 2.0492 & 2.0246 & 1.9205 & 2.0895 & 1.8758 \\
		5.5   & 2.2283 & 1.9098 & 2.1503 & 2.0412 & 2.0574 & 2.1043 & 2.1809 & 1.9562 & 2.0670 & 1.9620 & 2.2153 \\
		\hline
	\end{tabular}
	\label{tab: N and Tau with noise}
\end{table*}

\subsection{Influence of the Drag Coefficient}
\label{subsection: Influence of C_d}

\begin{figure}[hbt!]
	\centering
	\includegraphics[width=.45\textwidth]{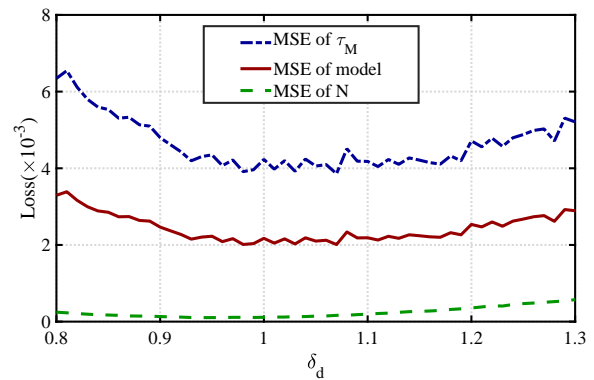}
	\caption{MSE related to $\delta_d$.}
	\label{fig: MSE of Cd}
\end{figure}

In this subsection, we present the performance of the model under different drag coefficients $C_d$. We enlarge or narrow the drag coefficient $\delta_d$ times and perform a 300-run Monte Carlo simulation for each condition. 
%The changes in MSE with $\delta_d$ increases is shown in \autoref{fig: MSE of Cd}. 
The changes of MSE are shown in \autoref{fig: MSE of Cd}. 
The figure shows that the identification accuracy of the model decreases with the deviation of the drag coefficient, and the influence of $\delta_d$ on the identification performance of $\tau_M$ is larger than that of $N$.
%Note that the deviation of $C_d$ affects on $\tau_M$ more than $N$.
This result occurs because the first-order lateral time constant $\tau_M$ is generally associated with drag coefficient $C_d$, so it is more sensitive to deviations in $C_d$.

\section{Conclusion}
\label{section: Conclusion}
% built a model
In this paper, a regression parameter identification model based on the gated neural network is established. The inputs of the model are available information between a aircraft and a missile, while the outputs are the guidance law parameter and the first-order lateral time constant. Numerical simulations verify the performance of the model and show that the guidance law parameter has a higher identification accuracy than the first-order lateral time constant.

% IMMM
To increase the training speed and the accuracy of the model, thought of transfer learning is introduced to MMM and an output processing method called improved multiple-model mechanism is proposed in this paper, which can be applied to general multiple-output regression problems. The main idea of IMMM is to set several groups of regimes in the multiple-model layer, with each group corresponding to a regression result of the neural network. A certain regression result is the weighted sum of regimes in the corresponding group. Simulation results show that compared with a conventional model, the model using IMMM has a faster training speed and a higher identification accuracy. Moreover, the model using IMMM can ensure the outputs of the model lie within a reasonable range.

\end{document}